\documentclass[]{llncs}
\usepackage{hetcasl}
\usepackage{amsmath}
\usepackage{amssymb}
\usepackage{alltt}
\usepackage{verbatim}

\renewcommand{\marginpar}[1]{}

\def\rhdsim{{\LARGE $\stackrel{\rhd}{\footnotesize \raisebox{-1ex}{$\sim$}}$} }

\begin{document}


\setcounter{secnumdepth}{7}

\newtheorem{hypothesis}{Hypothesis}

\newenvironment{preuve}{\noindent {\bf Proof}~}{\hfill$\Box$
\medskip
}

\newcommand{\Introduction}{\section*{Introduction}
                           \addcontentsline{toc}{chapter}{Introduction}}

\newtheorem{notation}{Notation}
\newtheorem{lemme}{Lemma}

\def\sl#1{\underline{#1}}
\def\cl#1{{\mathcal{#1}}}

\newcommand{\df}[1]{#1\!\!\searrow}
\newcommand{\udf}[1]{#1\!\!\nearrow}

\newtheorem{corollaire}{Corollary}

\newcommand{\rmq}{\hspace{-5mm}{\bf Remark:} ~}
\newcommand{\lpar}{\par\noindent}

\newenvironment{ex.}{
        \medskip
        \noindent {\bf Example}}
        {\hfill$\Box$ \medskip}
\newenvironment{Enonce}[1]{
        \begin{description}
        \item[{\bf #1~:}]
        \mbox{}
}{
        \end{description}
}

\def\Doubleunion#1#2#3{\displaystyle{\bigcup_{#1}^{#2} {#3}}}
\def\Union#1#2{\displaystyle{\bigcup_{#1} #2}}
\def\Intersect#1#2{\displaystyle{\bigcap_{#1} #2}}
\def\Coprod#1#2#3{\displaystyle{\coprod_{#1}^#2 #3}}
\def\Produit#1#2#3{\displaystyle{\prod_{#1}^{#2} {#3}}}
\def\Conj#1#2{\displaystyle{\bigwedge_{#1} \!\!\!\!\!#2}}

\newcommand{\Nat}{I\!\!N}
\newcommand{\Int}{Z\!\!\! Z}

\def\infer#1#2{\mbox{\large  ${#1} \frac {#2}$ \normalsize}}

\renewcommand{\marginpar}[1]{}

\def\rhdsim{{\LARGE $\stackrel{\rhd}{\footnotesize \raisebox{-1ex}{$\sim$}}$} }


\newcommand{\mv}{\varepsilon}

\def\M#1#2{M_{\scriptstyle #1 \atop \scriptstyle #2}}
\def\H#1#2{H_{\scriptstyle #1 \atop \scriptstyle #2}}
\newcommand{\Ro}{{\cal{R}}}
\newcommand{\Mo}{{\cal{M}}}
\newcommand{\Ho}{{\cal{H}}}
\newcommand{\Eo}{{\cal{E}}}
\newcommand{\ssucc}{\succ \!\! \succ}


\def\id#1{\underline{#1}}



\title{{\bf Testing Data Types Implementations from Algebraic Specifications}}
\author{Marie-Claude Gaudel$^1$, Pascale Le Gall$^2$}

\institute{$^1$ Universit\'e de Paris-Sud 11, LRI CNRS UMR 8623,\\ 
Bat. 490, F-91405 Orsay Cedex, France \\
{\tt mcg@lri.fr}\\
$^2$Universit\'e d'\'Evry-Val d'Essonne, IBISC CNRS FRE 2873,\\
523 pl. des Terrasses F-91025 \'Evry Cedex, France \\ 
{\tt pascale.legall@ibisc.univ-evry.fr} 
}

\maketitle
\date{February 3rd, 2008}

\begin{abstract}
  Algebraic specifications of data types provide a natural basis for
  testing data types implementations.  In this framework, the
  conformance relation is based on the satisfaction of axioms.
  This makes it possible to formally state the fundamental concepts of testing: exhaustive test set, testability hypotheses, oracle.\\
  \hspace*{0.2 cm} Various criteria for selecting finite test sets
  have been proposed. They depend on the form of the axioms, and on
  the possibilities of observation of the implementation under test. This last point is related to the well-known oracle problem. As the main interest of algebraic specifications is data type abstraction, testing a concrete implementation raises the issue of the gap between the abstract description and the concrete representation. The observational semantics of algebraic specifications bring solutions on the basis of the so-called observable contexts.\\
  \hspace*{0.2 cm} After a description of testing methods based on
  algebraic specifications, the chapter gives a brief presentation of
  some tools and case studies, and presents some applications to other
  formal methods involving datatypes.

\end{abstract}



\section{Introduction}
\label{intro} 

Deriving test cases from some descriptions of the Implementation Under
Test (the IUT) is a very old and popular idea.  In their
pioneering paper \cite{GG75}, Goodenough and Gerhart pointed out that
the choice of test cases should be based both on code coverage, and on
specifications expressed by condition tables.  One of the
first papers where software testing was based on some formal
description of the system under test, was by Chow \cite{Chow78}:
software was modelled by finite state machines. It has been very
influential on all the subsequent works on testing based on formal
specifications.

Most approaches in this area are based on behavioural descriptions:
for instance the control graph of the program, or some finite state
machine or labelled transition system.  In such cases, it is rather
natural to base the selection of test scenarios on some coverage
criteria of the underlying graph.

\medskip Algebraic specifications are different: abstract data types are
described in an axiomatic way \cite{Kreo99,CASLuserM,Wir90}.  There is
a signature $ \Sigma $, composed of a finite set $S$ of sorts  and
a finite set $F$ of function names over the sorts in $S$, and there
is a finite set of axioms $Ax$.  The correctness requirement is no
more, as above, the ability (or the impossibility) for the IUT to exhibit
certain behaviours: what is required by such specifications is the satisfaction of the axioms by the
implementation of the functions of $F$.  As a consequence, a natural
way for testing some IUT is to choose some instantiations of the axioms
(or of some consequences of them) and to check that when computed by
the IUT, the terms occurring in the instantiations yield results that
satisfy the corresponding axiom (or consequence).  This approach was
first proposed by Gannon et al. \cite{GMH81}, and
Boug\'e et al. \cite{Boug82,Boug86}, and then developed and
implemented by Bernot et al. \cite{BGM91}. 

Since these foundational works, testing from algebraic specifications
has been investigated a lot. Numerous works have addressed different
aspects.  

Some authors as in \cite{Barbey00} or
\cite{ClaessenH00} focus on a target programming language (Ada or
Haskell). 
Testing from algebraic specifications has also been
succesfully adapted for testing object-oriented systems \cite{TACCLE,DF94,Peraire98}.  
Besides, methods inspired
from algebraic testing have been applied to some other kinds of
specifications like model-based specifications, first by Dick et al.
\cite{DickFaivre93}, and more recently in \cite{DanAichernig04}. 
Some other works explore links
between test and proof
\cite{BerghoferN04,Bruckerwolff,Dydjer-HaiyanTakeyama03}.  

Some tools
\cite{Bruckerwolff,ClaessenH00,Mar95} based either on resolution
procedures or on specialised tactics in a proof engine, have been developed and used.  

Extensions of
algebraic specifications have also been studied, for instance,
bounded datatypes \cite{agm96} or partial functions \cite{al02}. More
recently, some contributions \cite{Doche-Wiels00,Machado00,MachadoS02}
have been done to take into account structured or modular
specifications aiming at defining structured test cases and at
modelling the activity of both unit testing and integration testing.

\medskip Another special feature of algebraic specifications is the abstraction gap between the abstract specification level and the concrete implementation. 
This raises problems for interpreting the results of  test experiments with respect to the specification.
This characteristic is shared with other formal methods that allow the description of complex datatypes in an abstract way, for instance VDM, Z, and their object oriented  extensions.

As a consequence, in the area of testing based on algebraic specifications, a special emphasis has been put on the oracle problem
\cite{al02,Bernot91,LeGall-Arnould96wadt,Machado98,Zhu03}.  The oracle
problem concerns the difficulty of defining reliable decision
procedures to compare values of terms computed by the IUT. Actually,
implementations of abstract data types may have subtle or complex
representations, and the interface of the concrete datatypes
is not systematically equipped with an equality procedure to compare values. In
practice, only some basic datatypes provide a reliable decision
procedure to compare values. They are said to be observable. The only
way to define (partial) decision procedure for abstract data types is
to observe them by applying  some (composition of) functions yielding
an observable result: they are called observable contexts.
Observational approaches of algebraic specifications bring solutions
to define an appropriate notion of correctness taking into account
observability issues.

The chapter is organised as follows: 
Section \ref{preliminaries} presents some necessary basic notions of algebraic specifications;
Section \ref{testing} gives the basic definitions of \emph{test} and
\emph{test experiment} against an algebraic specification; 
Section \ref{firstpres} introduces in a progressive way the notions of
\emph{exhaustive test set} and \emph{testability hypothesis} in a
simple case.  
Then Section \ref{select} addresses the issue of the
selection of a finite test set via the so-called \emph{uniformity} and
\emph{regularity} selection hypotheses.  
Section \ref{obs} develops
further the theory, addressing the case where there are observability
problems: this leads to a reformulation of the definitions mentioned
above, and to a careful examination of the notion of correctness. 
Section \ref{relw} presents some of the most significant related pieces of work.
The last section is devoted to brief presentations of some case studies,
and to the descriptions of some transpositions of the framework to
some other formal methods where it is possible to specify complex data
types.

\section{Preliminaries on algebraic specifications}
\label{preliminaries}

Algebraic specifications of data types, 
sometimes called axiomatic specifications,
provide a way of defining abstract data types by giving the properties (axioms) of
their operations.
There is no explicit definition of each operation (no pre- and post-condition, no algorithm)
but a global set of properties that describes the relationship between the operations.
This idea comes from the late seventies \cite{GTW78,GH78}. It has been the origin
of numerous pieces of work that have converged on the definition of 
\textsc{Casl}, the Common Algebraic Specification Language \cite{CASLuserM}. 

An example of an algebraic specification is given in Figure 1: it is a
\textsc{Casl} specification of containers of natural numbers, i.e. a data structure
that contains possibly duplicated numbers with no notion of order.
This specification states that there are three sorts of values, namely Natural Numbers, Booleans and Containers.
Among the operations, there is, for instance, a function named $isin$ which, given two values, resp. of sort
natural number and container, returns a boolean value. 
The operations must satisfy the axioms that are the formulas itemised by big bullets.

The sorts, operation names, and profiles of the operations are part of the 
{ \em signature}
of the specification. 
The signature gives the interface of the specified data type. Moreover, it declares some sorted variables that are used for writing the axioms.

An {\em (algebraic) signature} $\Sigma
= (S,F,V)$ consists of a set $S$ of sorts, a set $F$ of function names
each one equipped with an arity in $S^* \times S$ and a 
set of variables $V$, each of them being indexed by a sort. 
In the sequel, a function $f$ with arity
$(s_1 \ldots s_n,s)$, where  $s_1 \ldots s_n,s \in S$, will be noted $f:s_1 \times \ldots \times s_n
\rightarrow s$. 

In Figure 1,
the sorts of the signature are $Nat$ and $Bool$ (specified in some
\textsc{Our/Numbers/with/Bools} specification, not given here), and
$Container$; the functions are $[]$ (the empty container), $ \_ :: \_$
(addition of a number to a container), $isin$ that checks for the belonging 
of a number to a container, and $remove$ that takes away one occurrence of
a number from a container; the variables are $x$, $y$ of $Nat$ sort,
and $c$ of $Container$ sort.

\begin{figure}
\begin{hetcasl}
\FROM \=\SId{Our/Numbers/with/Bools} \VERSION 0.0 \GET \SId{Nat}, \SId{Bool}\\
\\
\SPEC \=\SIdIndex{Containers} \= \Ax{=}\\
\>\SId{Nat}, \SId{Bool}\\
\THEN\\
\>\GENERATEDTYPE \=\Id{Container} \Ax{::=}\=~[]~ \hspace*{-0.84mm}\textbar~\_\_\Ax{::}\_\_(\=\Id{Nat}; \Id{Container})~\\
\>\OP \=\Id{isin} :   \Id{Nat} \Ax{\times} \Id{Container} \Ax{\rightarrow} \Id{Bool}\\
\>\OP \=\Id{remove} :  \Id{Nat} \Ax{\times}  \Id{Container} \Ax{\rightarrow} \Id{Container}\\
\>\Ax{\forall} \Id{x}, \Id{y}: \Id{Nat}; \Id{c}: \Id{Container}\\
\>\Ax{\bullet} \= \Id{isin}(\=\Id{x}, []) \Ax{=} false \`{\small{}\KW{\%(}isin\_empty\KW{)\%}}\\
\>\Ax{\bullet} \= \Id{eq}(\= \Id{x}, \Id{y}) \Ax{=}  true \Ax{\Rightarrow} \Id{isin}(\=\Id{x}, \Id{y} \=\Ax{::} \Id{c}) \Ax{=} true \`{\small{}\KW{\%(}isin\_1\KW{)\%}}\\
\>\Ax{\bullet} \= \Id{eq}(\= \Id{x}, \Id{y}) \Ax{=}  false \Ax{\Rightarrow} \Id{isin}(\=\Id{x}, \Id{y} \=\Ax{::} \Id{c}) \Ax{=} \Id{isin}(\=\Id{x}, \Id{c})  \`{\small{}\KW{\%(}isin\_2\KW{)\%}}\\ 
\>\Ax{\bullet} \=\Id{remove}(\=\Id{x}, []) \Ax{=} [] \`{\small{}\KW{\%(}remove\_empty\KW{)\%}}\\
\>\Ax{\bullet} \=\Id{eq}(\= \Id{x}, \Id{y}) \Ax{=}  true \Ax{\Rightarrow} \Id{remove}(\=\Id{x}, \Id{y} \Ax{::} \Id{c}) \Ax{=} \Id{c} \`{\small{}\KW{\%(}remove\_1\KW{)\%}}\\

\>\Ax{\bullet} \=\Id{eq}(\= \Id{x}, \Id{y}) \Ax{=}  false \Ax{\Rightarrow} \Id{remove}(\=\Id{x}, \Id{y} \Ax{::} \Id{c}) \Ax{=} \Id{y} \Ax{::} \Id{remove}(\=\Id{x}, \Id{c}) \`{\small{}\KW{\%(}remove\_2\KW{)\%}}\\
\END\\
\end{hetcasl}
\caption{An algebraic specification of containers of natural numbers}
\end{figure}

Given a signature $\Sigma = (S,F,V)$, $T_\Sigma(V)$ is the set of {\em terms with variables
in $V$}  freely generated from variables and functions in
$\Sigma$ and preserving arity of functions. Such terms are indexed by the sort of their result.
We note $T_\Sigma(V)_s$ the subset of $T_\Sigma(V)$ containing exactly those terms indexed by $s$.

$T_\Sigma$ is the set $T_\Sigma(\emptyset)$ of the {\em ground terms} and we note $T_{\Sigma,s}$
the set of ground terms of sort $s$.


Considering the \textsc{Container} specification, an example of a ground term $t$ of $Container$ sort is $0  ::  0  ::  []$. An example of a term $t'$ with variables is $isin(x, 0  :: c)$ that is of $Bool$ sort.

A {\em substitution} is any mapping $\rho : V \rightarrow T_\Sigma(V)$ that preserves sorts. 
Substitutions are naturally extended to terms with variables.
The result of the application of a substitution $\rho$ to a term $t$ is called an {\em instantiation} of $t$, and is noted $t \rho$.
In the example, let us consider the substitution $\sigma : \{ x \rightarrow 0, y \rightarrow 0, c \rightarrow y::[] \}$, the instantiation $t' \sigma$ is the term with variable $isin(0, 0  ::  y  ::  [])$.

{\em $\Sigma$-equations} are formulae
of the form $t = t'$ with $t,t' \in
T_\Sigma(V)_s$ for $s \in S$. 
An example of an equation on containers is $remove(x,[]) = []$.

A {\em positive conditional $\Sigma$-formula} is any sentence of the
form $\alpha_1 \wedge \ldots \wedge \alpha_n \Rightarrow \alpha_{n+1}$
where each $\alpha_i$ is a $\Sigma$-equation ($1 \leq i \leq n+1$).
$Sen(\Sigma)$ is the set of all positive conditional
$\Sigma$-formulae.

A {\em (positive conditional) specification}  $SP = (\Sigma,Ax, C)$ consists of a signature $\Sigma$,
a set $Ax$ of positive conditional formulae often called {\em axioms}, and some 
{\em constraints} $C$, which may restrict the interpretations of the declared symbols (some examples are given below). 
When $C$ is empty, we note  $SP = (\Sigma,Ax)$ instead of  $SP = (\Sigma,Ax, \emptyset)$.

Specifications can be structured as seen in the example: a specification $SP$ can use some other specifications $SP_1, \ldots, SP_n$. 
In such cases, the signature is the union of signatures, and there are some {\em hierarchical} constraints that require the semantics of the used specifications to be preserved (for more explanations see \cite{Wir90}).

In the \textsc{Containers} specification, there are six axioms, named $isin\_empty$, $isin\_1$,  $isin\_2$, $remove\_empty$, $remove\_1$, and $remove\_2$, and there is a so-called {\em generation} constraint, expressed at the line beginning by 
\textbf{generated type}, that all the containers are computable by composition of the functions $[]$ and $ \_ :: \_$.
Such constraints are also called  {\em reachability constraints}.  
The functions $[]$ and $ \_ :: \_$ are called the {\em constructors} of the $Container$ type.

In some algebraic specification languages, axioms can be formulae of first-order logic, as in \textsc{Casl}. 
However, in this chapter we mainly consider positive conditional specifications\footnote{The reason is that most tools and case studies we present have been performed for and with this kind of specifications.
An extension of our approach to first order logic, with some restrictions on quantifiers, was proposed by Machado in \cite{Machado00}.}.

\medskip

A {\em $\Sigma$-algebra}
${\cal A}$ is
a family of sets $A_s$, each of them being indexed by a sort;
these sets are equipped, for each $f : s_1 \times \ldots \times
s_n \rightarrow s \in F$, with a mapping $f^{\cal A} : A_{s_1}
\times \ldots \times A_{s_n} \rightarrow A_s$. A $\Sigma$-morphism
$\mu$ from a $\Sigma$-algebra ${\cal A}$ to a $\Sigma$-algebra ${\cal
  B}$ is a mapping $\mu : A \rightarrow B$ such that for all $s
\in S$, $\mu(A_s) \subseteq B_s$ and for all $f:s_1 \times \ldots
\times s_n \rightarrow s \in F$ and all $(a_1,\ldots,a_n) \in A_{s_1}
\times \ldots \times A_{s_n}$ $\mu(f^{\cal A}(a_1,\ldots,a_n)) =
f^{\cal B}(\mu(a_1),\ldots,\mu(a_n))$. 

$Alg(\Sigma)$ is the class of all $\Sigma$-algebras. 

Intuitively speaking, an implementation of a specification with signature $\Sigma$ is a 
$\Sigma$-algebra: it means that it provides some sets of values named by the sorts,
and some way of computing the functions on these values without side effect. 

The set of ground terms
$T_\Sigma$ can be extended into a $\Sigma$-algebra by providing each
function name $f : s_1 \times \ldots \times s_n \rightarrow s \in F$
with an application $f^{T_\Sigma} : (t_1,\ldots,t_n) \mapsto
f(t_1,\ldots,t_n)$. In this case, the function names of the signature are simply interpreted
as the syntactic constructions of the ground terms.

Given a $\Sigma$-algebra ${\cal A}$, we note $\_^A
: T_\Sigma \rightarrow A$ the unique $\Sigma$-morphism that maps any
$f(t_1,\ldots,t_n)$ to $f^{\cal A}(t^A_1,\ldots,t^A_n)$. A
$\Sigma$-algebra ${\cal A}$ is said {\em reachable} if $\_^A$
is surjective.

A $\Sigma$-interpretation in $A$ is any
mapping $\iota : V \rightarrow A$. 
It is just an assignment of some values of the $\Sigma$-algebra to the
variables. Given such an interpretation, it 
is extended to terms with variables: the value of the term
is the result of its computation using the values of the variables and
the relevant $f^{\cal A}$.

A $\Sigma$-algebra ${\cal A}$ {\em
satisfies} a $\Sigma$-formula $\varphi:\wedge_{1 \leq i \leq n}~{t_i = t'_i}
\Rightarrow t = t'$, noted ${\cal A} \models \varphi$, if and only if for every $\Sigma$-interpretation
$\iota$ in $A$, if for all $i$ in $1..n$, $\iota(t_i) = \iota(t'_i)$ then $\iota(t) =
\iota(t')$. 
Given a specification $SP = (\Sigma,Ax, C)$, a $\Sigma$-algebra ${\cal
  A}$ is a {\em $SP$-algebra} if for every $\varphi \in Ax$, ${\cal A}
\models \varphi$ and ${\cal A}$ fulfils the $C$ constraint.  $Alg(SP)$
is the subclass of $Alg(\Sigma)$  exactly containing all the
$SP$-algebras. 

A $\Sigma$-formula $\varphi$ is a {\em semantic
  consequence} of a specification $SP = (\Sigma,Ax)$, noted $SP
\models \varphi$, if and only if for every $SP$-algebra ${\cal A}$, we
have ${\cal A} \models \varphi$.

\section{Testing against an algebraic specification}
\label{testing}

Let $SP$ be a positive conditional specification and $IUT$ be an Implementation Under Test. 
In dynamic testing, we are interested in the properties of the computations by $IUT$ of the functions specified in $SP$. 
$IUT$ provides some procedures or methods for executing these functions.
The question is whether they satisfy the axioms of $SP$. 

Given a ground $\Sigma$-term $t$, we note $t^{IUT}$ the result of its computation by $IUT$. 
Now we define how to test $IUT$ against a $\Sigma$-equation.

\begin{definition} [: Test and Test Experiment]
Given a $\Sigma$-equation $\epsilon$, and $IUT$ which provides an implementation for every function name of $\Sigma$, 
\begin{itemize}
\item
a test for $\epsilon$ is any ground instantiation $t = t'$ of $\epsilon$;
\item
a test experiment of $IUT$ against $t = t'$ consists in the evaluation of $t^{IUT}$  and $t'^{IUT}$  and the comparison of the resulting values.
\end{itemize}
\end{definition}

\begin{example}
One test of the $isin\_empty$ equation in the \textsc{Containers}
 specification of Figure 1 is $isin(0, []) = false$.
\end{example}

The generalization of this definition to positive conditional axioms is straightforward. 

In the following, we say that a test experiment is successful if it
concludes to the satisfaction of the test by the $IUT$, and we note it
$IUT \ passes \ \tau $ where $\tau$ is the test, i.e. a ground
formula. We generalise this notation to test sets: $IUT \ passes
  \ TS $ means that $\forall \tau \in TS, \ IUT \ passes \ \tau $.
  
Deciding whether $IUT \ passes \ \tau $ is the oracle problem
mentioned in the introduction.  In the above example it is just a
comparison between two boolean values.  However, such a comparison may
be difficult when the results to be compared have complex data types.
We postpone the discussion on the way it can be realised in such cases
to Section \ref{obs}.  Actually, we temporarily consider in the two
following sections that this decision is possible for all sorts, i.e.
they are all ``observable''.

\begin{remark} Strictly speaking, the definition above defines a
  tester rather than a test data: a test $t = t'$ is nothing else than
  the abstract definition of a program that evaluates $t$ and $t'$ via
  the relevant calls to the IUT and compares the results; a test
  experiment is an execution of this tester linked to the IUT.
\end{remark}

We can now introduce a first definition of an exhaustive test of an
implementation against an algebraic specification. A natural notion of
correctness, when all the data types of the specification are observable, is that the IUT
satisfies the axioms of the specification. Thus we start with a first
notion of exhaustive test inspired from the notion of satisfaction as
defined in Section \ref{preliminaries}.

\section{A first presentation of exhaustivity and testability}
\label{firstpres}

\begin{definition} [: Exhaustive Test Set, first version]
  Given a positive conditional specification $SP = (\Sigma , Ax)$, an exhaustive test set for $SP$, noted $Exhaust_{SP}$, is
  the set of all well-sorted ground instantiations of the axioms in
  $Ax$: $$
  Exhaust_{SP} = \{ \phi \rho \mid \phi \in Ax, \rho \in V
  \rightarrow T_{\Sigma} \} $$
\end{definition}

An exhaustive test experiment of some IUT against $SP$ is the set of
all the test experiments of the IUT against the formulas of
$Exhaust_{SP}$.

As said above, this definition is very close to (and is derived from)
the notion of satisfaction of a set of $\Sigma$-axioms by a
$\Sigma$-algebra.  In particular, the fact that each axiom can be
tested independently comes from this notion.

However, an implementation's passing once all the tests in the
exhaustive test set does not necessarily mean that it satisfies the
specification: first, this is true only if the IUT is deterministic;
second, considering all the well-sorted ground instantiations is, a
priori, not the same thing as considering all the
$\Sigma$-interpretations in the values of the IUT.  It may be the
case that some values are not expressible by ground terms of the
specification.

In other words, the above test set is exhaustive with respect to the
specification, but may be not with respect to the values used by the
program.  Thus some \emph{testability hypotheses} on the
implementation under test are necessary: the success of the exhaustive
test set ensures the satisfaction of the specification by the
implementation only if this implementation behaves as a reachable
$\Sigma$-algebra (cf. Section \ref{preliminaries}).

Practically, it means that: 
\begin{itemize}

\item There is a realisation of every function of $\Sigma$ that is supposed to be deterministic; the results do not depend on some hidden, non specified, internal state.

\item The implementation is assumed to be developed following good programming practices; any computed value of a data type must always be a result of the specified operations of this data type.

\item There is a comparison procedure for the values of every sort of the signature.

\end{itemize}

Note that, explicitly or not, all testing methods make
assumptions on IUT: a totally erratic system, or a diabolic one,
may pass some test set and fail later on\footnote{Testing methods
  based on Finite State Machine descriptions rely on the assumption
  that the IUT behaves as a FSM with the same number of states as
  the specification; similarly, methods based on IO-automata or
  IO-Transition Systems assume that the IUT behaves as an
  IO-automata: consequently, it is supposed input-enabled , i.e.
  always ready to accept any input.}.  In our case these hypotheses
are static properties of the program. Some of them are (or could be)
checkable by some preliminary static analysis of the source code.

\begin{definition} [: $\Sigma$-Testability]
  Given a signature $\Sigma$, an IUT is $\Sigma$-testable if it
  defines a reachable $\Sigma$-algebra ${\cal A}_{IUT}$. Moreover, for
  each $\tau$ of the form $t = t'$, there exists a way of deciding
  whether it passes or not.

The $\Sigma$-testability of the IUT is called the minimal hypothesis $H_{min}$ on the IUT.
\end{definition}

Let us note $Correct(IUT, SP)$ the correctness property that a given $IUT$
behaves as a reachable $SP$-algebra (i. e. the axioms are satisfied
and all the values are specified).  The fundamental link between
exhaustivity and testability is given by the following formula: $$
H_{min}(IUT) \Rightarrow ( \forall \tau \in Exhaust_{SP}, IUT \ passes
\ \tau \Leftrightarrow Correct(IUT, SP)) $$
$Exhaust_{SP}$ is
obviously not usable in practice since it is generally infinite.
Actually, the aim of the definitions of $Exhaust_{SP}$ and $H_{min}$
is to provide frameworks for developing theories of black-box testing
from algebraic specifications. Practical test criteria (i.e. those
which correspond to finite test sets) will be described as stronger
hypotheses on the implementation.  This point is developed in Sections
\ref{select} and \ref{obs}.

Before addressing the issue of the selection of finite test sets, let
us come back to the definition of $Exhaust_{SP}$.  As it is defined,
it may contain useless tests, namely those instantiations of
conditional axioms where the premises are false: such tests are
always successful, independently of the fact that their conclusion is
satisfied by the IUT or not. Thus they can be removed.

\begin{example}
  Assuming that $eq(0,0)=true$ is a semantic consequence of the 
  \textsc{Our/Numbers/with/Bools} specification, we can derive an equational test for the $remove\_1$
  conditional axiom in the \textsc{Containers}
 specification of Figure 1.  This test is simply the ground equation:\\
 $remove(0, 0 :: 0 :: []) = 0 :: []$.
\end{example}

In the example of Figure 1, we have distinguished a subset of
functions as constructors of the $Container$ type (namely $[]$ and
$::$).  Under some conditions, the presence of constructors in a
specification makes it possible to characterise an equational
exhaustive test set.

A {\em signature with constructors} is a signature $\Sigma = <S, F,V>$
such that a subset $\cal{C}$ of elements of $F$ are distinguished as
constructors.  Let us note $\Omega = <S, \cal{C}$, $V>$ the
corresponding sub-signature of $\Sigma$, and $T_\Omega$ the
corresponding ground terms.  A specification $SP = <\Sigma, Ax>$ where
$\Sigma$ is a signature with constructors $\cal{C}$ is {\em complete
  with respect to its constructors} if and only if both following
conditions hold:

\begin{itemize}
\item $\forall t \in T_\Sigma, \exists t' \in T_\Omega$ such that $SP \models t = t'$
\item $\forall t, t' \in T_\Omega, SP \models t = t' \Rightarrow \  <\Sigma, \emptyset> \models t = t' $, i.e. $t$ and $t'$ are syntactically identical
\end{itemize}

\begin{example}
  The \textsc{Containers}
 specification of Figure 1 is complete with
  respect to the constructors $\cal{C} = \{[], :: \}$ of the
  $Container$ sort: from the axioms, any ground term of $Container$
  sort containing some occurrence of the (non constructor) $remove$
  function is equal to some ground term containing only occurrences of
  $[]$ and $::$. Moreover, there is only one such ground term.
\end{example}

For such specifications and under some new hypotheses on the IUT, it
is possible to demonstrate that the set of ground conclusions of the
axioms is exhaustive. When removing premises satisfied
by the specification, we should be careful not to remove some other
premises that the IUT could interpret as true, even if they are not
consequences of the specification. A sufficient condition is to
suppose that the IUT correctly implements the constructors of
all the sorts occurring in the premises. Let us introduce the new testability hypothesis
$H_{min,{\cal C}}$ for that purpose. Intuitively, $H_{min,{\cal C}}$
means that the IUT implements data types with a syntax very close to
their abstract denotation. It may seem to be a strong hypothesis, but in
fact, it only applies to basic types, often those provided by the implementation language.
As soon as the data type implementation is subtle or complex,
the data type is then encapsulated and thus considered as non
observable for testing (cf. Section \ref{obs}).

\begin{definition}
$IUT$ satisfies $H_{min,{\cal C}}$ iff $IUT$ satisfies $H_{min}$ and :

$$\forall s \in S,
\forall u,v \in T_{\Omega,s}, \;  IUT \; passes \; u = v \Leftrightarrow SP
\models u = v$$
\end{definition}

\begin{definition}
$$
\begin{array}{ll}
EqExhaust_{SP,\cal{C}} = \{ & \epsilon \rho  \mid  
\exists \alpha_1 \wedge \ldots \wedge
\alpha_n \Rightarrow \epsilon \in Ax, 
\\
& 
\rho \in V \rightarrow T_{\Omega}, 
SP \models (\alpha_1 \wedge \ldots \wedge
\alpha_n)\rho \}
\end{array}
$$
\end{definition}

Under $H_{min,{\cal C}}$ and for specifications complete with
respect to their constructors $EqExhaust_{SP,\cal{C}}$
is an exhaustive test set. 
A proof can be found in \cite{LeGall93} or in \cite{AABLM05}. 
Its advantage over $Exhaust_{SP}$ is that it is made of equations. 
Thus the test experiments are simpler.

Some other approaches for the definitions of exhaustivity and
testability are possible. For instance, as suggested in \cite{BGM92}
and applied by Dong and Frankl in the ASTOOT system \cite{DF94}, a
different possibility is to consider the algebraic specification as a
term rewriting system, following a ``normal-form'' operational
semantics.  Under the condition that the specification defines a
ground-convergent rewriting system, it leads to an alternative
definition of the exhaustive test set: $$Exhaust'_{SP} = \{ t =
t\downarrow \mid t \in T_\Sigma \}$$
where $ t\downarrow$ is the
unique normal form of $t$. The testability hypothesis can be weakened
to the assumption that the IUT is deterministic (it does not need
anymore to be reachable).  In \cite{DF94}, an even bigger exhaustive
test set was mentioned (but not used), which contained for every
ground term the inequalities with other normal forms, strictly
following the definition of initial semantics.

Actually, this is an example of a case where the exhaustive test set is not built from instantiations of the axioms, but more generally from an adequate set of semantic consequences of the specification. Other examples are shown in Section \ref{obs}. 

\section{Selection hypotheses: uniformity, regularity}
\label{select}

\subsection{Introduction to selection hypotheses}
\label{hypIntro}

A black-box testing strategy can be formalised as the selection of a
finite subset of some exhaustive test set. In the sequel, we work with
$EqExhaust_{SP,\cal{C}}$, but what we say is general to the numerous
possible variants of exhaustive test sets.

Let us consider, for instance, the classical partition testing strategy\footnote{more exactly, it should be called sub-domain testing strategy.}. 
It consists in defining a finite collection of (possibly non-disjoint) subsets that covers the exhaustive test set. 
Then one  element of each subset is selected and submitted to the implementation under test.
The choice of such a strategy corresponds to stronger hypotheses than $H_{min}$ on the implementation under test. 
We call such hypotheses {\em selection hypotheses}. 
In the case of partition testing, they are called {\em uniformity hypothesis}, since
the implementation under test is assumed to uniformly behave on some test subsets $UTS_i$ (as Uniformity Test Subset):

$$
UTS_1 \cup \ldots \cup UTS_p = EqExhaust_{SP,\cal{C}}, \mbox{ and }
$$
$$
\forall i = 1, \ldots, p,
( \forall \tau \in UTS_i, IUT \ passes \ \tau \Rightarrow IUT \ passes \ UTS_i )
$$

Various selection hypotheses can be formulated and combined depending
on some knowledge of the program, some coverage criteria of the
specification and ultimately cost considerations.  Another type of
selection hypothesis is {\em regularity hypothesis}, which uses a size
function on the tests and has the form ``if the subset of
$EqExhaust_{SP,\cal{C}}$ made up of all the tests of size less than or
equal to a given limit is passed, then $EqExhaust_{SP,\cal{C}}$ also
is"\footnote{\label{regul}As noticed by several authors, \cite{DF94},
\cite{Chen-Tse-Chan-Chan98}, and from our own experience \cite{MTW92},
such hypotheses must be used with care. It is often necessary to
choose this limit taking in consideration some ``white-box knowledge"
on the implementation of the datatypes: array bounds, etc}.

All these hypotheses are important from a theoretical point of view because they formalise common test practices and express the gap between the success of a test strategy and correctness. 
They are also important in practice because exposing them makes clear the assumptions made on the implementation. 
Thus, they give some indication of complementary verifications, as used by Tse et al. in \cite{Chen-Tse-Chan-Chan98}. 
Moreover, as pointed out by Hierons in \cite{Hier02}, they provide formal bases to express and compare test criteria and fault models. 

\subsection{How to choose selection hypotheses}

As said above, the choice of the selection hypotheses may depend on
many factors.  However, in the case of algebraic specifications, the
text of the specification provides useful guidelines.  These
guidelines rely on coverage of the axioms and composition of the cases
occurring in premise of the axioms via unfolding as stated first in \cite{BGM91},
and extended recently in \cite{AABLM05}.

We recall that axioms are of the form
$\alpha_1 \wedge \ldots \wedge \alpha_n \Rightarrow \alpha_{n+1}$ where each $\alpha_i$ is a
$\Sigma$-equation $t_i = t'_i$, ($1 \leq i \leq n+1$).

From the definition of $EqExhaust_{SP,\cal{C}}$, a test of such an axiom is
some $\alpha_{n+1} \rho$ where $\rho \in V \rightarrow T_{\Sigma}$
is a well-typed ground substitution of the variables of the axiom such
that the premise of the axiom, instantiated by
$\rho$, is true: it is a semantic consequence of the specification
($SP \models (\alpha_1 \wedge \ldots \wedge \alpha_n)\rho$).

One natural basic testing strategy is to cover each axiom once, i. e. to choose for every axiom one adequate substitution $\rho$ only. The corresponding uniformity hypothesis is

$
 \forall \rho \in V \rightarrow T_{\Sigma} \ such \ that \ SP \models (\alpha_1 \wedge \ldots \wedge \alpha_n)\rho, \ IUT \ passes \ \alpha_{n+1}\rho \Rightarrow$\\
$ (IUT \ passes \ \alpha_{n+1}\rho',
\forall \rho'\in V \rightarrow T_{\Sigma} \ such \ that \ SP \models (\alpha_1 \wedge \ldots \wedge \alpha_n)\rho' )
$

It defines a so-called {\em uniformity sub-domain} for the variables
of the axiom that is the set of ground $\Sigma$-terms
characterised by $SP \models (\alpha_1 \wedge \ldots \wedge
\alpha_n)$.

\begin{example}
  In the example of Figure 1, covering the six axioms
  requires six tests, for instance the following six ground equations:
\begin{itemize}
\item $isin(0,[]) = false$, with the whole $Nat$ sort as uniformity sub-domain; 
\item $isin(1, 1::2::[]) = true$, with the pairs of $Nat$ such that $eq(x, y) = true$ and the whole $Container$ sort as uniformity sub-domain; 
\item $isin(1, 0::3::[]) = isin(1, 3::[])$, with the pairs of $Nat$ such that $eq(x, y) = false$ and the whole $Container$ sort as uniformity sub-domain;
\item $remove(1, []) = []$, with the $Nat$ sort as uniformity sub-domain;
\item $remove(0, 0::1::[]) = 1::[]$, with the pairs of $Nat$ such that  $eq(x, y) = true$ and the $Container$ sort as uniformity sub-domain; 
\item $remove(1, 0::[]) = 0::remove(1,[])$, with the pairs of $Nat$ such that $eq(x, y) = false$ and the $Container$ sort as uniformity sub-domain. 
\end{itemize}
\end{example}

Such uniformity hypotheses are often too strong. A method for
weakening them, and getting more test cases, is to compose the cases occurring in the axioms. In the
full general case, it may involve tricky pattern matching on the
premises and conclusions, and even some theorem proving. However,
when the axioms are in a suitable form one can use the classical
unfolding technique defined by Burstall and Darlington in \cite{BD77}.
It consists in replacing a function call by its definition. Thus, for
unfolding to be applicable, the axioms must be organised as a set of
functions definitions: every function is defined by a list of
conditional equations such as:

$\wedge_{1 \leq i \leq m}~{\alpha_i} \Rightarrow f(t_1,\ldots,t_n) = t$

where the domain of the function must be covered by the disjunction of the premises of the list.

\begin{example}
In the example of Figure 1, the $isin$ function is defined by:
\begin{hetcasl}
\>\Ax{\bullet} \= \Id{isin}(\=\Id{x}, []) \Ax{=} false \`{\small{}\KW{\%(}isin\_empty\KW{)\%}}\\
\>\Ax{\bullet} \= \Id{eq}(\= \Id{x}, \Id{y}) \Ax{=}  true \Ax{\Rightarrow} \Id{isin}(\=\Id{x}, \Id{y} \=\Ax{::} \Id{c}) \Ax{=} true \`{\small{}\KW{\%(}isin\_1\KW{)\%}}\\
\>\Ax{\bullet} \= \Id{eq}(\= \Id{x}, \Id{y}) \Ax{=}  false \Ax{\Rightarrow} \Id{isin}(\=\Id{x}, \Id{y} \=\Ax{::} \Id{c}) \Ax{=} \Id{isin}(\=\Id{x}, \Id{c})  \`{\small{}\KW{\%(}isin\_2\KW{)\%}}\\ 
\end{hetcasl}
It means that every occurrence of $isin(t_1, t_2)$ can correspond to the three following sub-cases:
\begin{itemize}
\item 
$t_2 = []$: in this case $isin(t_1, t_2)$ can be replaced by $false$;
\item
$t_2 = y :: c$ and $eq(t_1, y) = true$: in this case, it can be replaced by $true$;
\item
$t_2 = y :: c$ and $eq(t_1, y) = false$: in this case, it can be replaced by $y :: isin(t1, c)$.
\end{itemize}
\end{example}
A way of partitioning the uniformity sub-domain induced by the
coverage of an axiom with some occurrence of $f(t_1,\ldots,t_n) = t$
is to introduce the sub-cases stated by the definition of $f$, and, of
course, to perform the corresponding replacements in the conclusion
equation to be tested.  This leads to a weakening of the uniformity
hypotheses.

\begin{example}
  Let us consider the $isin\_2$ axiom. Its coverage corresponds to the
  uniformity sub-domain ``pairs of $Nat$ such that $eq(x, y) = false$''
 $\times$ ``the $Container$ sort". Let us unfold in this axiom the second
  occurrence of $isin$, i.e. $isin(x, c)$. It leads to three 
  sub-cases for this axiom:
\begin{itemize}
\item
$c = []$: \\
$eq(x, y) = false \wedge c = []  \Rightarrow isin(x, y :: []) = isin(x,[])$, i.e, false;
\item
$c = y' :: c'$ and $ eq(x, y') = true$ : \\
$eq(x, y) = false \wedge c = y' :: c' \wedge eq(x, y') = true \Rightarrow isin(x, y :: y' :: c') = isin(x, y' :: c')$, i.e., true;
\item
$c = y' :: c'$ and $ eq(x, y') = false$ : \\
$eq(x, y) = false \wedge c = y' :: c' \wedge eq(x, y') = false \Rightarrow isin(x, y :: y' :: c') = y :: isin(x, y' :: c')$, i.e. $isin(x, c')$.
\end{itemize}
The previous uniformity sub-domain is partitioned in three smaller
sub-domains characterised by the three premises above. Covering these sub-cases leads to
test bigger containers, and to check that $isin$ correctly behaves
independently of the fact that the searched number was the last to be
added to the container or not. Applying the same technique to the
$remove\_2$ axiom leads to test that in case of duplicates, one
occurrence only is removed.
\end{example}

Of course, unfolding can be iterated: the last case above can be
decomposed again into three sub-cases.  Unbounded unfolding leads
generally to infinite test sets\footnote{Actually, as it is described
  here, unbounded unfolding yields an infinite set of equations very
  close to the exhaustive test set. The only remaining variables are
  those that are operands of functions without definitions, namely, in
  our case, constructors}.  Limiting the number of unfoldings is
generally sufficient for ensuring the finiteness of the test set.
Experience has shown (see Section \ref{casestudies}) that in practice
one or two levels of unfolding are sufficient for ensuring what test
engineers consider as a good coverage and a very good detection power.
In some rare cases, this limitation of unfolding does not suffice for
getting a finite test set: then, it must be combined with regularity
hypotheses, i. e. limitation of the size of the ground instantiations.


\medskip Unfolding has been implemented by Marre within the tool
LOFT~\cite{BGM91,Mar91,Mar95} using logic programming. 
There are
some conditions on the specifications manipulated by LOFT:
\begin{itemize}
\item they must be complete with respect to constructors;  
\item when transforming the specification into a conditional rewriting system (by
orienting each equation $t = t'$ occuring in an axiom from left to
right $t \rightarrow t'$), the resulting conditional rewrite system must be
confluent and terminating;
\item each equation $t =t'$ that is
the conclusion of an axiom must be such that $t$ may be decomposed as
a function $f$, not belonging to the set of constructors, applied to a
tuple of terms built on constructors and variables only.  
\end{itemize}
Under these
conditions, the LOFT tool can decompose any uniformity domain
into a family of uniformity sub-domains. It can also compute some
solutions into a given uniformity sub-domain. These two steps
correspond respectively to the computation of the uniformity hypotheses based
on unfolding subdomains and to the generation of an arbitrary test case per
each computed subdomain. The unfolding procedure is based on an
equational resolution procedure involving some unification mechanisms.
Under the conditions on the specifications given above, the
unfolding procedure computes test cases such that: sub-domains are
included in the domain they are issued from (soundness), and the
decomposition into subdomains covers the splitted domain
(completeness). 

In \cite{AABLM05}, Aiguier et al. have extended
the unfolding procedure for positive conditional specifications without restrictions.
This procedure is also sound and complete.
However, the price to pay is that instead of unfolding a unique occurrence of
a defined function, the extended unfolding procedure requires to
unfold all occurrences of the defined functions in a given equation
among all the equations characterising the domain under decomposition.
This may result in numerous test cases.

\medskip We have seen that conditional tests can be simplified into equational ones by solving their premises.
It can be done in another way, replacing variables
occurring in the axiom by terms as many times as necessary to find
good instantiations.  This method amounts to draw terms as long as the
premises are not satisfied. This is particularly adapted in a
probabilistic setting. In \cite{BernotBG97}, Bouaziz et al. give
some means to build some distributions on the sets of values.

\section{Exhaustivity and testability versus observability}
\label{obs}

Until now, we have supposed that a test experiment $t=t'$ of the IUT
may be successful or not depending on whether the evaluations of $t$
and $t'$ yield the same resulting values. Sometimes,
comparing the test outputs may be a complex task when some information
is missing. It often corresponds to complex abstract data types
encapsulating some internal concrete data representations. Some
abstract data types (sets, stacks, containers, etc) do not always provide an equality
procedure within the implementation under test and we reasonnably
cannot suppose the existence of a finite procedure,
the oracle, to correctly interpret the test results as equalities or
inequalities.  The so-called {\em oracle problem} in the framework of
testing from algebraic specifications amounts to deal with equalities
between terms of non observable sorts.

In this section, we distinguish a subset $S_{Obs}$ of observable sorts
among the set $S$ of all sorts. For example, it may regroup all the
sorts equipped with an equality predicate within the IUT
environnement, for instance equality predicates provided by the
programming language and considered as reliable. The minimal
hypothesis $H_{min}$ is relaxed to the weaker hypothesis
$H_{min}^{Obs}$ expressing that the the IUT still defines a
reachable $\Sigma$-algebra but that the only remaining elementary
tests which may be interpreted by the IUT as a verdict
success/failure are the ground equality $t = t'$ of observable sort.
The set $Obs$ of all observable formulae is the subset of
$Sen(\Sigma)$ of all formulae built over observable ground equalities.
Any formula of $Obs$ may be considered as a test experiment,
and reciprocally.

The oracle problem in the case of non observable sorts may be tackled by
two distinct but related questions. How to turn non observable
equalities under test into test experiments tractable by an $IUT$ only
satisfying $H_{min}^{Obs}$? How far can we still talk about correctness
when dealing with observability issues? Roughly speaking, the
answers lie  respectively in using observable contexts and in defining
correctness up to some observability notion. We present these two
corresponding key points
in the following sections.

\subsection{Observable contexts}
\label{obs-context}

In practice, non observable abstract data types can be observed
through successive applications of functions leading to an observable
result. It means that properties related to non observable sorts can
be tested through observable contexts :

\begin{definition} [: Context and Observable context]

An observable context $c$
  for a sort $s$ is a term of observable sort with a unique occurrence
  of a special variable of sort $s$, generically denoted by
  $z$. 

Such a context is often denoted
  $c[z]$ or simply $c[.]$ and $c[t]$ denotes $c\sigma$
  where $\sigma$ is the substitution associating the term $t$ to the
  variable $z$.

An observable context is said to be minimal if it does contain an
observable context as a strict subterm\footnote{A subterm of a term $t$ is $t$ itselt or any term occurring in it. In particular, if $t$ is of form $f(t_1,\ldots,t_n)$ then $t_1$, \ldots and $t_n$ are subterms of $t$. A strict subterm of $t$ is any subterm of $t$ which differs from $t$.}.
\end{definition}

Only minimal observable
  contexts are meaningful for testing. Indeed, if a context $c$ has an
  observable context $c'$ as a strict subterm, then $c[z]$ may be
  decomposed as $c_0[c'[z]]$. It implies that for any terms $t$
  and $t'$, $c[t]=c[t']$ iff $c'[t]=c'[t']$. Both equalities being
  observable, the simpler one, $c'[t]=c'[t']$, suffices to infer
  whether $c[t]=c[t']$ holds or not. In the sequel, all the
  observable contexts will be considered as minimal by default.

For example, we can use set cardinality and element membership to
observe some set data type as well as the height and the top of all
successive popped stacks for some stack data type. Thus, a non
observable ground equality of the form $t=t'$ is observed through all
observable contexts $c[.]$ applied to both $t$ and $t'$.  From a
testing point of view, it amounts to apply to both terms $t$ and $t'$
the same successive application of operations yielding an observable
value, and to compare the resulting values.

\begin{example}
With the \textsc{Containers}
specification of Figure 1, we now consider that the sort $Container$ is no more an
observable sort while $Nat$ and $Bool$ are observable ones. Ground equalities of
sort $Container$ should be observed through the observable sorts $Nat$
and $Bool$. An
abstract test like $remove(3,[]) = []$ is now observed through
observable contexts. Each observable context of sort $Container$ gives
rise to a new (observable) test belonging by construction to $Obs$.
For example, the context $isin(3, z)$ applied to the previous abstract
test leads to the test : $isin(3, remove(3,[])) = isin(3,[])$. 
\end{example}

In practice, there is often an infinity of such observable contexts. 
In the case of the \textsc{Containers}
specification, we can build the following
observable contexts\footnote{For convenience, we use the variables
  $x$, $x_1$, \ldots, $x_n$ to denote arbitrary ground terms of sort
  $Nat$ in a concise way.} 

\begin{center}
{\small $isin(x, x_1 :: ( x_2 :: \ldots (x_n ::
z)))$, $isin(x, remove(x_1,remove(x_2,\ldots,remove(x_n,z))))$}
\end{center}

 or more generally, any combination of the operations $remove$ and $::$
surrounded by the $isin$ operation. 
As a consequence, we are facing a new kind of selection problem: to test an equality
$t=t'$ of $Container$ sort, one has to select among all these
observable contexts a subset of finite or even reasonable size.

Bernot in \cite{Bernot91} gives a counter-example based on the stack
data type to assess that without additional information on the IUT,
all the contexts are a priori necessary to test a non observable
equality, even those involving constructors such as $::$. More
precisely, a context of the form $isin(x, x_1 :: z)$ may appear
useless since it leads to build larger $Container$ terms
instead of observing the terms replacing $z$. In \cite{Bernot91}, it is shown
that those contexts may reveal some programming errors
depending on a bad use of state variables. From a theoretical point of
view, let us consider a specification reduced to one axiom $a = b$
expressing that two non observable constants are equal. Then for any
given arbitrary minimal context $c_0$, one can design a program
$P_{c_0}$ making  $c[a]=c[b]$ true for all minimal observable contexts
except $c_0$. This fact means that in general, any minimal context is
needed to ``fully'' test non observable equalities. This is a
simplified explanation of a proof given by Chen,Tse et al. in
\cite{Chen-Tse-Chan-Chan98}.

  Let us point out that replacing an equation $t = t'$ by the
  (infinite) set of $c[t]=c[t']$ with $c$ an observable context is 
  classical within the community of algebraic
  specifications.  Different observational
  approaches~\cite{Bidoit-Hennicker-Wirsing95,ONS91} have been
  proposed to cope with refinement of specifications based on abstract
  data types.  They have introduced the so-called behavioural
  equalities, denoted by $t \approx t'$. The abstract equality is
  replaced by the (infinite) set of all observables contexts applying
  to both terms. More precisely, an algebra ${\cal A}$ satisfies $t
  \approx t'$ if and only if for every $\Sigma$-interpretations
  $\iota$ in $A$, for all observable contexts $c$, we have
  $\iota(c[t])=\iota(c[t'])$. Behavioural equalities allow the
  specifier to refine abstract data types with concrete data types
  that do not satisfy some properties required at the abstract level.
  For example, the $Set$ abstract data type with some axioms
  stating the commutativity of the element insertion, can be refined
  into the $List$ abstract data type where the addition of an element by
  construction cannot be commutative. The refinement of $Set$ by $List$
  is ensured by requiring that equalities on sets hold in the list
  specification only up to the behavioural equality. It amounts to
  state that  observable operations (here the membership operation)
  behave in the same way at the abstract level of sets and at the
  implementation level of lists and to ignore those properties of the
  implementation that are not observable.

  Considering an infinity of contexts is possible using context
  induction as defined by Hennicker in \cite{Hennicker91}. This is
  useful to prove a refinement step, but is useless in order to define
  an oracle.  So, how can we select a finite set of observable
  contexts? Below we give some hints:
\begin{itemize}
\item The selection hypotheses presented in Section~\ref{select} to choose particular
  instantiations of axiom variables can be transposed to choose observable
  contexts. In particular, a rather natural way of
  selecting contexts consists in applying a regularity hypothesis. The
  size of a context is often defined in relation with the number of
  occurrences of non observable functions occurring in it.
\item If one can characterise the equality predicate by means of a set
  of axioms, then one can use this axiomatisation, as proposed by
  Bidoit and Hennicker in \cite{Bidoit-Hennicker96}, to define the
  test of non observable equalities. To give an intuition of how such
  an axiomatisation looks like, we give below the most classical one.
  It concerns the specification of abstract data types like sets, bags
  or containers, for which two terms are equal if and only if they
  exactly contain the same elements. Such an axiomatisation looks like: 
   \begin{center}
  $c \approx c'$ iff  $\forall e, isin(e,c) = isin(e,c')$
   \end{center}
  where $c$ and $c'$ are variables of the abstract data type to be
  axiomatised, and $e$ is a variable of element sort. $c \approx c'$
  denotes the behavioural equality that is axiomatised. The
  axiomatisation simply expresses that the subset of contexts of the
  form $isin(e,z)$ suffices to characterize the behavioural equality.
  This particular subset of contexts can then be chosen as a suitable
  starting point to select observable contexts to test non observable
  equalities. Such an approach has two main drawbacks. First, such a
  finite aximatisation may not exist\footnote{For example, the
    classical stack specification has no finite axiomatisation of
    stack equality.}  or be difficult to guess. Second, selecting only
  from the subset of observable contexts corresponding to a finite
  axiomatisation amounts to make an additional hypothesis on the
  IUT, which has been called the {\em oracle hypothesis} in
  \cite{Bernot91}. In a few words, it consists in supposing that the
  IUT correctly implements the data type with respect to the
  functions involved in the axiomatisation. In the example of Containers,
  two containers are supposed to be behaviourally equal if and only if the
  membership operation $isin$ applied on the containers always gives the same
  results. In other words, by using axiomatisation to build oracles,
  we are exactly supposing what we are supposed to test.  Clearly, it
  may appear as a too strong hypothesis.
\item Chen, Tse and others in \cite{Chen-Tse-Chan-Chan98} point out
  that some static analysis of the IUT may help to choose an adequate
  subset of observable contexts. When testing whether $t=t'$ holds or
  not, the authors compare their internal representations $r$ and $r'$
  within the IUT. If $r$ and $r'$ are equal, then they can
  conclude\footnote{\cite{Machado98} is partially based on this same
    idea : if the concrete implementations are identical, then
    necessarily their corresponding abstract denotations are equal
    terms.} that the IUT passes $t=t'$. Otherwise, if $r$ and $r'$ are
  not equal, then they study which data representation components are
  different in $r$ and $r'$ and which are the observations which may
  reveal the difference. This makes it possible to build a subset of
  observable contexts which has a good chance to observationally
  distinguishes $t$ and $t'$. The heuristic they have proposed has
  been successfully applied in an industrial context
  \cite{Tse-Lau-Chan-Liu-Luk06}.
 \end{itemize}

\subsection{Correctness with observability issues}
\label{correct-obs}

We have seen in Section~\ref{obs-context} that the test of a non
observable equality may be approached by a finite subset of observable
contexts. More precisely, a non observable ground equality $t=t'$ may be
partially verified by submitting a finite subset of the test set:
$$Obs(t=t') = \{c[t]=c[t'] \mid c \mbox{ is a minimal observable context}
\}$$.

The next question concerns testability
issues : can we adapt the notions of correctness and exhaustivity 
when dealing with observability ?  For example, one may wonder whether the set
$Obs(t=t')$ may be considered as an exhaustive test set for testing
the non observable (ground) equality $t=t'$. More generally, by taking
inspiration from the presentation given in
Section~\ref{firstpres}, we look for a general property linking
the notions of exhaustive test set and  testability such as:

$$
H_{min}^{Obs}(IUT) \Rightarrow ( \forall \tau \in Exhaust_{SP}^{Obs},
IUT \ passes \  \tau \Leftrightarrow Correct^{Obs}(IUT, SP))
$$

\subsubsection{Equational specifications} \mbox{~} \\

If $SP$ is an equational specification\footnote{Axioms of an equational
specification  are of the form $t=t'$ where $t$ and $t'$ are terms with
  variables and of the same sort.}, then following
Section~\ref{obs-context}, the test set

$$
\begin{array}{ll}
Exhaust_{SP}^{Obs} = \{ & c[t] \rho = c[t'] \rho   \mid  t = t' \in Ax,
\rho \in V \rightarrow T_\Sigma, \\
&  c \mbox{ minimal  observable context} \}
\end{array}
$$

is a good candidate\footnote{Let us remark that if $t$ and $t'$ are of
  observable sort $s$, then the only minimal observable context is
  $z_s$ such that $t \rho=t' \rho$ are the unique tests associated to
  the axiom $t=t'$. } since it simply extends the $Obs(t=t')$ sets to
the case of equations with variables. Actually, $Exhaust_{SP}^{Obs}$
is an exhaustive test set provided that we reconsider the
definition of correctness taking into account observability.

By definition of observability, the IUT does not give access to any information on non observable
sorts.  Considering a given $IUT$ as correct with respect to some
specification $SP$ should be defined up to all the possible
observations and by discarding properties directly expressed on non
observable sorts. Actually, observational correctness may be
defined as : $IUT$ is observationally correct with respect to $SP$
according to the set of observations $Obs$, if there exists an
$SP$-algebra ${\cal A}$ such that $IUT$ and ${\cal A}$ exactly
behave in the same way for all possible observations. 

To illustrate,
let us consider the case of the Container specification enriched by a new axiom of  commutativity of  element insertions:
$$x::(y::c)=y::(x::c)$$ 
The $Container$ datatype is classically
implemented by the $List$ data type. However, elements in lists are
usually stored according to the order of their insertion. In
fact, the $List$ data type is observationally equivalent to the $Container$
data type as soon as the membership element is correctly implemented
in the $List$ specification. It is of little matter whether the $List$
insertion function satisfies or not the axioms concerning the addition
of elements in Containers.

This is formalised by introducing
equivalence relations between algebras defined up to a set of
$\Sigma$-formulae.

\begin{definition}

Let  $\Psi \subset Sen(\Sigma)$ and  ${\cal A}$ and ${\cal
  B}$ be two $\Sigma$-algebras. 

${\cal A}$ is said to be $\Psi$-equivalent to ${\cal B}$, denoted
by ${\cal A} \equiv_{\Psi} {\cal B}$, if and only if we have $\forall
\varphi \in \Psi, {\cal A} \models \varphi \Longleftrightarrow {\cal B}
\models \varphi$.
 
${\cal A}$ is said to be observationally equivalent to ${\cal B}$ if
and if ${\cal A} \equiv_{Obs} {\cal B}$.

\end{definition}

We can now define observational correctness:

\begin{definition}
 
Let $IUT$ be an implementation under test satisfying $H_{min}^{Obs}$.

$IUT$ is observationally correct with respect to  $SP$ and according
to $Obs$, denoted by $Correct^{Obs}(IUT,SP)$ if and only if

{\em $$\exists {\cal A} \mbox{ reachable } SP \mbox{-algebra}, IUT
\equiv_{Obs} {\cal A}$$
}
\end{definition}

\begin{remark}
  This notion of observational correctness has been first recommended
  for testing purpose by Le Gall and Arnould in
  \cite{LeGall93,LeGall-Arnould96wadt} for a large classe of
  specifications and observations\footnote{For interested readers,
    \cite{BGM91,LeGall93,LeGall-Arnould96wadt} give a generic
    presentation of formal testing from algebraic specifications in
    terms of institutions.}. With respect to the observational
  approaches in algebraic specifications
  \cite{Bidoit-Hennicker-Wirsing95}, it corresponds to abstractor
  specifications for which the set of algebras is defined as the set
  of all algebras equivalent to at least an algebra of a kernel set,
  basically the set of all algebras satisfying the set of axioms.
\end{remark}

From a testing point of view, each reachable $SP$-specification is
obviously observationally correct with respect to $SP$. Reciprocally,
an implementation $IUT$ is observationally correct if it cannot be
distinguished by observations from at least a reachable $SP$-algebra,
say $IUT_{SP}$. So, nobody can say whether the implementation is the
$SP$-algebra $IUT_{SP}$, and thus intrinsically correct, or the $IUT$
is just an approximation of one reachable $\Sigma$-algebra up to the
observations $Obs$.  Thus, under the hypothesis $H_{min}^{Obs}$, any
observationally correct $IUT$ should be kept. Finally,
$Correct^{Obs}(IUT,SP)$ captures  exactly the set of all implementations
which look like $SP$-algebras up to the observations in $Obs$.  With
this appropriate definition of $Correct^{Obs}(IUT,SP)$, the test set
$Exhaust_{SP}^{Obs}$ is exhaustive. A sketch of the proof is the
following. For each $IUT$ passing $Exhaust_{SP}^{Obs}$, let us
consider the quotient algebra ${\cal Q}$ built from $IUT$ with the
axioms of $SP$.  We can then show that ${\cal Q}$ is a
$SP$-algebra and is observationally equivalent to $IUT$.

\subsubsection{Positive conditional specifications with observable
  premises} \mbox{~} \\

We also get an exhaustive test set when considering
axioms with observable premises. For each axiom of the form $\alpha_1 \wedge \ldots \wedge
\alpha_n \Rightarrow t = t'$ with all $\alpha_i$ of observable sort,
it suffices to put in the corresponding exhaustive test set all the
tests of the form $\alpha_1 \rho \wedge \ldots \wedge
\alpha_n \rho \Rightarrow c[t] \rho = c[t'] \rho$ for all
substitutions $\rho : V \rightarrow T_{\Sigma}$ and for all minimal
observable contexts $c$.

Moreover, if we want to have an exhaustive test set involving equations
only, as it has been done in Section~\ref{firstpres}, we should restrict
to specifications with observable premises and complete with respect
to the set ${\cal C}_{Obs}$ of constructors of observable sorts. As in
Section~\ref{firstpres}, we also consider that the IUT correctly
implements the constructors of all the sorts occurring in the premise,
here the observable sorts\footnote{When observable sorts coincide with
the basic data types of the programming language, such an hypothesis
is quite plausible. Thus, this is a weak hypothesis}. That is to
say, $IUT$ satisfies $H_{min,{\cal C}_{Obs}}$ iff $IUT$ satisfies
$H_{min}$ and :

$$\forall s \in S_{Obs},
\forall u,v \in T_{\Omega,s} IUT \; passes \; u = v \Leftrightarrow SP
\models u = v$$

Under $H_{min,{\cal C}_{Obs}}$ and for the considered restricted class
of specifications (i.e. observable premises and completeness with
respect to ${\cal C}_{Obs}$), 

$$
 \begin{array}{ll}
EqExhaust_{SP}^{Obs} = & \{ c[t] \rho = c[t'] \rho   \mid  \exists \alpha_1 \wedge \ldots \wedge
\alpha_n \Rightarrow t = t' \in Ax,
\rho \in V \rightarrow T_\Sigma, \\
&
~~ c \mbox{ min.
  obs. context}, 
SP \models (\alpha_1 \wedge \ldots \wedge
\alpha_n)\rho \}
\end{array}
$$
is an exhaustive test set with respect to observational
correctness. 

\subsubsection{Generalisation to non-observable premises} \mbox{~} \\

Is it possible to generalise such a construction of an exhaustive test
set for specifications with positive conditional formulas comprising
non-observable premises? A first naive solution would consist in
replacing each non-observable equation $t=t'$ occuring either in the
premise or in the conclusion of the axioms by a subset of
$Obs(t=t')$. Unfortunately, such an idea cannot be applied, unless one
accepts to submit biased tests\footnote{A test is said to be biased
  when it rejects at least a correct implementation.}. This fact has
been reported by Bernot and others in \cite{Bernot91,BGM91}. To give
an intuition, let us consider a new axiom $$
x :: x :: l = x :: l
\Rightarrow true = false $$
which means that if addition to a
container is idempotent, then\footnote{This is no more than a positive conditional
  way of specifying $ x :: x :: l \neq x :: l$. Actually, as the
  trivial algebra (with one element per sort) is satisfying all the
  conditional positive specifications, the inconsistency of
  specifications is often expressed by the possibility of deriving the
  boolean equation $true=false$.} it would lead to $true=false$.  Let
us try to test the ground instance $ 0 :: 0 :: [] = 0 :: []
\Rightarrow true = false$ by considering a test $\phi$ in $Obs$ of the
form $$\Conj{ \small \begin{array}{l} \psi_i \in Obs(0 :: 0 :: [] = 0
    ::[]) \\ i \in I, I \mbox { finite index}
\end{array}}{\psi_i} \Rightarrow u = v$$ 
then the $IUT$ may pass the premise 
$$
\Conj{ \small \begin{array}{l}
    \psi_i \in
  Obs(0 :: 0 :: [] = 0 ::[]) \\ i \in I, I \mbox { finite index}
\end{array}}{\psi_i}
$$ 
without $0 :: 0 :: [] = 0 :: 0 :: []$ being
a consequence of the specification. In that case, the IUT passes
the test $\phi$ by passing the conclusion $true = false$. Thus, observing non
observable premises through a finite set of contexts leads to require
an observable equality, here $true=false$, which in fact is not required by the
specification. This is clearly a bad idea. 

It is now widely recognised
that non-observable equations may be observed through some subset of
observable contexts only when their position in the test is
positive\footnote{ Roughly speaking, an atom $t=t'$ is said to be in a positive
  position if by
  putting the test into disjunctive normal form, then the $t=t'$ is not
  preceded by a negation.}. For example, the disjunctive normal form 
of  $ 0 :: 0 :: [] = 0 :: []
\Rightarrow true = false$ is $\neg (0 :: 0 :: [] = 0 :: []) \vee true
= false$ and thus $ 0 :: 0 :: [] = 0 :: []$ has a negative position in
the test. In particular, Machado in \cite{Machado98,Machado00} considers any first
order formula whose Skolem form does not contain existential
quantifiers. Every non-observable equations in positive positions are
observed by means of observable contexts while those in negative
positions are observed by using concrete equality in the
implementation. In that sense, Machado's approach is not a
pure black-box approach deriving test cases and oracles from
specifications but an approach mixing black-box and white-box where
test cases are derived from the specifications  and the oracle
procedure is built from both the specification and the IUT.

We have shown that to deal with axioms with non-observable
premises, it is not possible to apply observable contexts.
However, can we do something else to handle such axioms? A tempting
solution is to use the specification to recognise some ground
instances of the axiom for which the specification requires the non
observable premise to be true.

Let us come back to the axiom $$
x :: x :: l = x :: l \Rightarrow true
= false$$

If it stands alone, nothing  can be done to test it.  Let us introduce
a new axiom stating the idempotence law on the element
insertion: 
$$ eq(x,y) = true \Rightarrow x :: y :: l = y :: l$$ Any ground
instance of $x :: x :: l = x :: l$ is then a semantic consequence of
the specification such that $true = false$ also becomes a semantic
consequence. In such a case, one would like to consider $true = false$
as a test and even more, it seems rather crucial to precisely submit
this test! This small example illustrates clearly why in this case,
tests cannot be only ground instances of axioms but shoud be selected
among all the observable semantic consequences of the
specification\footnote{Observable semantic consequences are just those
semantic consequences that belong to $Obs$. By construction, selecting
a test outside this set would reject at least one correct
implementation.} (see the end of Section~\ref{preliminaries} for the
definition of semantic consequence.).  Let us remark that according to
the form of the specifications, one can use the unfolding technique
described in Section~\ref{select} in order to solve the premise in the
specification. In \cite{LeGall93}, Le Gall has shown that when the
specification is complete with respect to the set ${\cal C}_{Obs}$ of
constructors of observable sorts and under $H_{min,{\cal C}_{Obs}}$,

$$
 \begin{array}{ll}
EqExhaust_{SP}^{Obs} = & \{ c[t] \rho = c[t'] \rho   \mid  \exists \alpha_1 \wedge \ldots \wedge
\alpha_n \Rightarrow t = t' \in Ax,
\rho \in V \rightarrow T_\Sigma, \\
&
~~ c \mbox{ min.
  obs. context}, 
SP \models (\alpha_1 \wedge \ldots \wedge
\alpha_n)\rho \}
\end{array}
$$
is an exhaustive test set with respect to observational
correctness. Curiously, whether there are non-observable premises or
not in the specification, the corresponding equational exhaustive test set is not
modified.

\section{Related work}
\label{relw}

\subsection{Related work on selection}

In
\cite{ClaessenH00}, Claessen and Hughes propose the QuickCheck tool for
randomly testing Haskell programs from algebraic specifications.
Axioms are encoded into executable Haskell programs whose arguments
denote axiom variables.  Conditional properties are tested by drawing
data until finding a number, given as parameter, of cases which
satisfy the premises. Of course, the procedure is stopped when a too
large number of values is reached. The QuickCheck tool provides the
user with test case generation functions for any arbitrary Haskell
datatype, and in particular, also for functional types. The user can
observe how the random data are distributed over the datatype
carrier.  When he considers that the distribution is not well balanced
on the whole domain, for instance if the premises are satisfied by
data of small size only, it is possible to specialise the test
case generation functions to increase the likelihood to draw values ensuring a
better coverage of the domain of premise satisfaction. This last
feature is very useful for dealing with dependent datatypes.  In
\cite{BerghoferN04}, Berghofer and Nipkow use Quickcheck to exhibit
counter-examples for universally quantified formulae written in
executable Isabelle/HOL.  This is a simple way to rapidly debug
formalisation of a theory.  In \cite{Dydjer-HaiyanTakeyama03}, Dydjer
et al. develop a similar approach of using functional testing
technics to help the proof construction by analysing counter-examples.

In \cite{Bruckerwolff}, Brucker and Wolff use the
full theorem proving environment Isabelle/HOL to present a method and
a tool HOL-TestGen for generating test cases. They recommand to take
benefit of the Isabelle/HOL proof engine equipped with tactics to
transform a test domain (denoted as some proof goal) into test subdomains
(denoted as proof subgoals). Selection hypotheses are expressed as
proof hypotheses and the user can interact to guide the test data
generation.  Both the Quickcheck and TestGen tools present the advantage
of offering an unified framework to deal with the specification, the
selection and the generation of test cases, and even the submission of
the test cases and the computation of the test verdict.

\subsection{Related work on observability}

We have given a brief account of observability considerations and
their important impact on testability issues.  In particular, there does
not always exist an exhaustive test set, since such an existence
depends on some properties of the specification and the
implementation: namely, restrictions on the specification and
hypotheses on the implementation.

The importance of observability issues for the oracle problem as been first 
raised by Bougé \cite{Boug86} and then Bernot, Gaudel
and Marre in \cite{Bernot91,BGM91}. It has been studied later on
by Le Gall and Arnould \cite{LeGall-Arnould96wadt} and Machado
\cite{al02,Machado98}.  
Depending on the hypotheses on the
possible observations and on the form or the extensions of the
considered specifications, the oracle problem has been specialised.
For example, in \cite{agm96}, Arnould et al. define a framework
for testing from specifications of bounded data types. To some
extent, bounds of data types limit the possible observations : any
data out of the scope of the bound description should not be observed
when testing against such specifications.  The set of
observable formulae are formulae which are observable in the classical
sense, where all terms are computed as being under the specified bound. 

As soon as partial function are considered in the specification, it
must be observable whether a term is defined or not.  In \cite{al02},
Arnould and Le Gall consider specifications with partial functions
where definedness can be specified using an unary predicate $def$. The
specification of equalities are declined with two predicates, strong
equality $=$ allowing two undefined terms to be equal; existential
equality $\stackrel{e}{=}$ for which only defined terms may be
considered as equal. As the predicate $\stackrel{e}{=}$ may be
expressed in term of $=$ and $def$, testing from specification with
partiality naturally introduces two kinds of elementary tests directly
related to the predicates $def$ and $=$.  Testing with partial
functions requires to take into account the definition predicate:
intuitively, testing whether a term is defined or not systematically
precedes the following testing step, that is testing about equality of
terms.  Some initial results about testability and exhaustive test sets
can be found in
\cite{al02}.

\subsection{Variants of exhaustivity}

Most exhaustive test sets presented here are made of tests directly
derived from the axioms: tests are ground instances of (conclusions
of) axioms, some equalities being possibly surrounded by observable
contexts. Such tests do not necessarily reflect the practice of
testing. 
Actually, the usual way of testing consists in applying the
operation under test to some tuples of ground constructor terms and to
compare the value computed by the IUT to a ground contructor term
denoting the expected result. This can be described by tests of the
form: $$f (u_1, \ldots, u_n) =v$$
with $f$ the function to be tested,
and $u_1,\ldots,u_n,v$ ground constructor terms. The underlying
intuition is that the constructor terms can denote all the
concrete values manipulated by the implementation (reachability constraint). 
To illustrate this
point of view, in the case of the \textsc{Containers}
specification and by
considering again that the sort $Container$ is observable, for the
axiom $remove\_2$, instead of testing $remove(2,3::[])=3::remove(2,[])$
by solving the premise $eq(2,3)=false$, a test of the good form would
be $remove(2,3::[])=3::[]$. Such a test may be obtained by applying
the $remove\_1$ axiom to the occurrence $remove(2,[])$.  In
particular, LOFT \cite{Mar91,Mar95} computes tests of this reduced
form. In \cite{AABLM05,al02,agm96}, Arnould et al. present some
exhaustive tests built from such tests involving constructor terms as
much as possible. 

\subsection{The case of structured specifications}
Until now, we have considered flat specifications which consist
of a signature, a set of axioms, and possibly reachability constraints. 
Moreover, we have studied the distinction between
observable and non observable sorts. Observable sorts often
correspond to the basic types provided by the
programming environnement, and non observable sorts to the type of interest for the
specification. 
However, algebraic specifications may be structured using various
primitives allowing to import, combine, enrich, rename or forget
(pieces of) imported specifications. Such constructions should be taken into
account when testing.

As a first step to integration
testing of systems described by structured algebraic specifications, Machado in
\cite{Machado00,MachadoS02} shows how to build a test set whose structure is guided by the structure of
the specification. 
The main and significant drawback of this
approach is that hidden operations are ignored. As soon as an
axiom involves an hidden operation, the axiom is not tested.
Depending on the organisation of the specification, this can mean that a lot of properties
are removed from the set of properties to be tested.

In \cite{Doche-Wiels00}, Doche and Wiels define a framework for composing test cases
according to the structure of the specification.  Their approach
may be considered as modular since the IUT should have the same
structure as the specification and the tests related to the
subspecifications are composed together. These authors have established that
correctness is preserved under some hypotheses\footnote{For interested
  readers, the hypotheses aim at preserving properties along signature
  morphisms and thus, are very close to the satisfaction condition of
  the institution framework.} and have applied their approach to an
industrial case study reported in \cite{Doche-Seguin-Wiels99}.

\section{Case studies and applications to other formal methods}
\label{casestudies}

This part of the paper briefly reports some case studies and experiments related to the theory presented here. 
Some of them were performed at LRI, some of them elsewhere. 
The first subsection is devoted to studies based on algebraic specifications. 
The next one reports interesting attempts to transpose some aspects of the theory to other formal approaches, namely VDM, Lustre, extended state machines and labelled transition systems. 
A special subsection presents some applications to object-oriented descriptions.

\subsection{First case studies with algebraic specifications}

A first experiment, performed at LRI by Dauchy and Marre, was on the on-board part of the driving system of an automatic subway\footnote{precisely, the train controller on line D in Lyon that has been operating since 1991.} in collaboration with a certification agency. 
An algebraic specification was written \cite{DauOz91}.
Then two critical modules of the specification were used for experiments with LOFT: the overspeed controller and the door opening controller.
These two modules shared the use of eight other specification modules that described the state of the on-board system. 
The number of axioms for the door controller was 25, with rather complex premisses. 
The number of axioms of the speed controller was 34.
There where 108 function names and several hundred axioms in the shared modules.
Different choices of uniformity hypotheses were experienced for the door controller: they led to 230, 95, and 47 tests. For the overspeed controller, only one choice was sensible and led to 95 tests.
The experiment is reported in details in  \cite{DGM93}.
In a few words, these tests were used by the certification team as a sort of checklist against the tests performed by the development team. This approach led to the identification of a tricky combination of conditions that had not been tested by the developers.

A second experiment is reported in \cite{MTW92} 
and was performed within a collaboration between LRI and the LAAS laboratory in Toulouse. 
The experiment was performed on a rather small piece of software written in C, which was extracted from a nuclear safety shutdown system.
The piece of software contained some already known bugs that were discovered but one: it was related to some hidden shared variable in the implementation, and required rather large instantiations, larger than the bound chosen a priori for the regularity hypothesis. 
On a theoretical point of view, this can be analysed as a case where the testability hypothesis was not ensured. 
More practically, the fault was easy to detect by ``white-box" methods, either static analysis or structural testing with branch coverage. 
This is coherent with the remark in Section \ref{firstpres} on the possibility of static checking of the testability hypothesis, and with the footnote \ref{regul} in Section  \ref{select} on the difficulties to determine adequate bounds for regularity hypotheses.

An experiment of ``intensive" testing of the EPFL library of Ada components was led by Buchs and Barbey in the Software Engineering Laboratory at EPFL \cite{Barbey00}.
First an algebraic specification of the component was reengineered: the signature was derived from the package specifications of the family, and the axioms were written manually. Then the LOFT system was used with a standard choice of hypotheses.

LOFT has been also used for the validation of a transit node algebraic specification \cite{Transit}. Generating test cases was used for enumerating scenarios with a given pattern. 
It led to the identification of one undesirable, and unexpected, scenario in the formal specification.

It was also used for the test of the data types of an implementation of the Two-Phase-Commit protocol \cite{GJ98} without finding any fault: this was probably due to the fact that the implementation had been systematically derived from a formal specification. Other aspects of this case study are reported in the next subsection.

The specifications and test sets of these case studies are too large to be given here. Details can be found in \cite{DauOz91} and  \cite{DGM93} for the first one, in  \cite{Transit} and \cite{TSI} for the transit node, and 
in  \cite{JEG99} for the Two-Phase-Commit protocol.

\subsection{Applications to other methods}

Actually, the approach developed here for algebraic data types is rather generic and
presents a general framework for test data selection from formal specifications.
It has been reused for, or has inspired, several test generation methods from various specification formalisms: VDM, Lustre, full LOTOS.

The foundational paper by Jeremy Dick and Alain Faivre on test case generation from VDM specifications
 \cite{DickFaivre93}  makes numerous references to some of the notions and techniques presented here, namely uniformity and regularity hypotheses, and unfolding.
The formulae of VDM specifications are relations on states decribed by operations (in the sense of VDM, i.e. state modifications). 
 They are expressed in first-order predicate calculus. 
 These relations are reduced to a disjunctive normal form (DNF), creating a set of disjoint sub-relations. Each sub-relation yields a set of constraints which describe a single test domain. 
The reduction to DNF is similar to axiom unfolding: uniformity and regularity hypotheses appear in relation with this partition analysis. 
As VDM is state-based, it is not enough to partition the operations domains. Thus the authors give a method of extracting a finite state automaton from a specification. 
This method uses the results of the partition analysis of the operations to perform a partition analysis of the states. 
This led to a set of disjoint classes of states, each of which corresponds either to a precondition or a postcondition of one of the above subrelations. 
Thus, a finite state automaton can be defined, where the states are some equivalence classes of states of the specifications. 
From this automaton, some test suites are produced such that they ensure a certain coverage of the automaton paths. 
The notion of test suites is strongly related to the state orientation of the specification: it is necessary to test the state evolution in presence of sequences of data, the order being important.

Test generation from Lustre descriptions has been first studied
jointly at CEA and LRI. The use of the LOFT system to assist the test of Lustre programs has been investigated. 
Lustre is a description language for reactive systems which is based on the synchronous approach
\cite{halbwachs91}. 
An algebraic semantics of Lustre was stated and entered as a specification in LOFT. 
Lustre programs were considered as enrichments of this specification, just as some specific axiom to be tested. 
After this first experience, GATEL, a specific tool for Lustre was developed by Marre at CEA (Commissariat \`a l'\'Energie Atomique). 
In GATEL, a Lustre specification of the IUT, and  some Lustre descriptions of environment constraints and test purpose are interpreted via Constraint Logic Programming. 
Unfolding is the basic technique, coupled with a specific constraint solving library \cite{Gatel,Marre-Blanc04}. 
GATEL is used at IRSN (Institut de Radioprotection et S\^uret\'e Nucl\'eaire) for identifying those reachable classes of tests covering a given specification, according to some required coverage criteria.
The functional tests performed by the developers are then compared to these classes in order to point out uncovered classes, i.e. insufficient testing. 
If it is the case, GATEL provides test scenarios for the missing classes.

LOTOS is a well known formal specification language, mainly used in the area of communication protocols.
There are two variants: basic LOTOS makes it possible to describe processes and their synchronisation, with no notion of data type; full LOTOS, where it is possible to specify algebraic data types and how their values can be communicated or shared between specified processes.
In the first case, the underlying semantics of a basic LOTOS specification is a finite labelled transition system. 
There is an extremely rich corpus of testing methods based on such finite models (see \cite{2} for an annotated bibliography). 
However, there are few results on extending them to infinite models, as it is the case when non trivial data types are introduced. In \cite{GJ98}, Gaudel and James have stated the underlying notions of testability hypotheses, exhaustive test sets, and selection hypotheses for full LOTOS.

This approach has been used by James for testing an implementation of the Two-Phase-Commit Protocol developed from a LOTOS specification into Concert/C. 
The results of this experiment are reported in \cite{JEG99}.
As said in the previous sub-section, tests for the data types were obtained first with the LOFT system. 
Then a set of testers was derived manually from the process part of the specification. 
The submission of these tests, was preceded by a test campaign of the implementations of the atomic actions of the specification by the Concert/C library, i. e. the communication infrastructure (the set of gates connecting the processes), which was developed step by step. 
It was motivated by the testability hypothesis: it was a way of ensuring the fact that the actions in the implementation were the same as in the specification, and that they were atomic.
No errors were found in the data types implementations, but an undocumented error of the Concert/C pre-processor was detected when testing them. 
Some errors were discovered in the implementation of the main process. 
They were related to memory management, and to the treatment of the time-outs. 
There are always questions on the interest of testing pieces of software, which have been formally specified and almost directly derived from the specification. 
But this experiment shows that problems may arise: the first error-prone aspect, memory management, was not expressed in the LOTOS specification because of its abstract nature; 
the second one was specified in a tricky way due to the absence of explicit time in classical LOTOS. Such unspecified aspects are unavoidable when developing efficient implementation.

\subsection{Applications to object-oriented software}

It is well known that there is a strong relationship between abstract
data types and object orientation.  There is the same underlying idea
of encapsulation of the concrete implementation of data types.  Thus
it is not surprising that the testing methods presented here for
algebraic specifications has been adapted to the test of object
oriented systems. We present two examples of such adaptations.

The ASTOOT approach was developped by Dong and Frankl at the
Polytechnic University in New-York \cite{DF94}.  The addressed problem
was the test of object-oriented programs: classes are tested against
algebraic specifications.  A set of tools had been developed. As
mentioned at the end of Section \ref{firstpres}, a different choice
was made for the exhaustive test set, which is the set of equalities
of every ground term with its normal form, and it was also suggested
to test inequalities of ground terms As normal forms are central in
the definition of tests, there was a requirement that the axioms of
the specification must define a convergent term rewriting system.
Moreover, there is a restriction to classes such that their operations
have no side effects on their parameters and functions have no side
effects: it corresponds to a notion of testability.  The oracle
problem was addressed by introducing a notion of observational
equivalence between objects of user-defined classes, which is based on
minimal observational contexts, and by approximating it.  Similarly to
Section \ref{select}, the test case selection was guided by an
analysis of the conditions occuring in the axioms; the result was a
set of constraints that was solved manually.  The theory presented
here for algebraic data types turned out to nicely fit to cope with
object-orientation, even when different basic choices were made.

This had been confirmed by further developments by Tse and its group
at the university of Hong Kong
\cite{Chen-Tse-Chan-Chan98,TACCLE,Tse-Lau-Chan-Liu-Luk06}.  In their approach,
object-oriented systems are described by algebraic specifications for
classes and contract specification for clusters of related classes :
contracts specify interactions between objects via message-passing
rules.  As in our approach, some tests are fundamental pairs of
equivalent ground terms obtained via instantiations of the axioms. As
in ASTOOT non equivalent pairs of terms are also considered.  Some
white-box heuristic for selecting relevant observable contexts makes
it possible to determine whether the objects resulting from executing
such test cases are observationally equivalent.  Moreover, message
passing test sequences are derived from the contract specification and
the source code of the methods.  This method has been recently applied
for testing object-oriented industrial software
\cite{Tse-Lau-Chan-Liu-Luk06}.

\section{Conclusion}

Algebraic specifications have proved to be an interesting basis for stating some
theory of black-box testing and for developing methods and tools.
The underlying ideas have turned out to be rather general and applicable to specification methods including datatypes, whatever the formalism used for their description. 
It is the case of the notions of uniformity hypothesis, and regularity hypotheses that have been reused in other contexts.

In presence of abstraction and encapsulation, the oracle problem
raises difficult issues due to the limitations on the way concrete
implementations can be observed and interpreted.  This is not specific
to algebraic specifications and abstract data types: the same problems
arise for embedded and/or distributed systems. It is interesting to
note the similarity between the observable contexts presented here,
and the various ways of distinguishing and identifying the state
reached after a test sequence in finite state machines \cite{LeeY96},
namely separating families, distinguishing sequences, characterising
sets, and their variants.

The methodology presented here has been applied, as such or with some
adjustments, in a significant number of academic or industrial case
studies.  In most cases, they have been used for some a posteriori
certification of critical systems that had already been intensively
validated and verified, or for testing implementations that have been
developed from some formal specification.  This is not surprising: in
the first case, the risks are such that certification agencies are
ready to explore sophisticated methods; in the second case, the
availability of the formal specification pushes for using it for test
generation.  In both circumstances, it was rather unlikely to find
errors.  But some were discovered however, and missing test cases were
identified.  In some cases, these detections were quite welcome and
prevented serious problems.  This is an indication of the interest of
test methods based on formal specifications, and of the role they can
play in the validation and verification process.

\bibliographystyle{plain}
\bibliography{biblioSurvey}



\end{document}